\documentclass[preprint,showpacs,preprintnumbers,amsmath,amssymb]{revtex4}
\usepackage{epsf}
\usepackage{color}
\usepackage{dcolumn}% Align table columns on decimal point
\usepackage{bm}% bold math
\everymath{\displaystyle}

\begin{document} %%%%%%%%%%%%%%%%%%%%%%%%%%%%%%%%%%%%%%%%%%%%%%%%%

\title{Spin dynamics in gravitational fields
of rotating bodies and the equivalence principle}

\author{\firstname{Yuri N.}~\surname{Obukhov}}
\email{yo@thp.Uni-Koeln.DE} \affiliation{Department of Mathematics,
University College London, Gower Street, London, WC1E 6BT, UK}

\author{\firstname{Alexander J.}~\surname{Silenko}}
\email{silenko@inp.minsk.by} \affiliation{Research Institute of
Nuclear Problems, Belarusian State University, Minsk 220080,
Belarus}

\author{\firstname{Oleg V.}~\surname{Teryaev}}
\email{teryaev@thsun1.jinr.ru} \affiliation{Bogoliubov Laboratory
of Theoretical Physics, Joint Institute for Nuclear Research,
Dubna 141980, Russia}
%\date{file ``ostfin12.tex", \today}

\begin{abstract}
We discuss the quantum and classical dynamics of a particle with spin
in the gravitational field of a rotating source. A relativistic equation
describing the motion of classical spin in curved spacetimes is obtained.
We demonstrate that the precession of the classical spin is in a perfect
agreement with the motion of the quantum spin derived from the Foldy-Wouthuysen
approach for the Dirac particle in a curved spacetime. We show that the
precession effect depends crucially on the choice of a tetrad. The results
obtained are compared to the earlier computations for different tetrad
gauges.
\end{abstract}

\pacs{04.20.Cv; 04.80.Cc; 03.65.Sq}
\maketitle

%%%%%%%%%%%%%%%%%%%%%%%%%%%%%%%%%%%%%%%%%%%%%%%%%
\section{Introduction}
%%%%%%%%%%%%%%%%%%%%%%%%%%%%%%%%%%%%%%%%%%%%%%%%%

A rotation of a central body, defining a difference between
stationary and static spacetimes, leads to an appearance of
specific gravitational effects. The most important effect has been
predicted by Lense and Thirring \cite{LT}. It consists in frame
dragging around rotating bodies and is manifested in a precession of
satellite orbits and gyroscopes (i.e. classical spins). The
nonrelativistic formula for the latter effect has been derived by
L. Schiff \cite{Schiff} (and further refined and generalized in
\cite{Shiff2}).

In the present work, we analyze quantum and classical spins in
stationary spacetimes. We use the weak-field approximation when all
components of the metric tensor $g_{ij}$ are close to the corresponding
components of the Minkowski tensor $\eta_{ij}$ ($|h_{ij}|\equiv|g_{ij}
-\eta_{ij}|\ll1$). The formulas calculated for classical spins extend
the previously obtained results to the relativistic case. The
investigation of the quantum dynamics of spins is carried out
for the first time.

The theory of classical spin in the first-order (linear) approximation
can be formulated as follows. A particle is characterized by its position
in spacetime, $x^i(\tau)$ which is a function of the proper time $\tau$,
and by the 4-vector of spin $S^\alpha$. The 4-velocity of a particle
$U^\alpha = e^\alpha_i dx^i/d\tau$ is normalized by the condition
$g_{\alpha\beta}U^\alpha U^\beta = c^2$ where $g_{\alpha\beta} = {\rm diag}
(c^2, -1, -1, -1)$ is the flat Minkowski metric. In order to be able to
describe spinning particles both in flat and curved spacetime (as well as
in arbitrary curvilinear coordinates), we use the tetrad $e^\alpha_i$ to
transform the components of different objects from the coordinate basis
(associated with the local coordinates $x^i$) to a local orthonormal
frame. When the gravitational field is absent, it is possible to choose
the Cartesian coordinates everywhere and use the holonomic orthonormal frame
which coincides with the natural frame, so that $e^\alpha_i = \delta^\alpha_i$
then. In general, the tetrad coefficients satisfy $g_{\alpha\beta}e^\alpha_i
e^\beta_i = g_{ij}$ for an arbitrary spacetime metric $g_{ij}$.

The foundations of the classical theory of particles with spin were laid down
by Mathisson and Papapetrou \cite{Mathisson,Papapetrou} (for a review see
\cite{corben}, e.g.).
Pomeransky, Khriplovich \cite{PK} and Dvornikov \cite{Dvornikov} developed
the relativistic approach for the equation of motion defining the dynamics
of three-component physical spin in curved spacetimes. This equation perfectly
describes the dynamics of the spin in static spacetimes. Here we present a
rigorous deduction of the equation of motion of the three-component spin,
confirming the heuristic arguments of \cite{PK}. At the same time, we show
here that this equation can be used for nonstatic spacetimes only with the
special choice of tetrads satisfying the condition $e^{0}_{\widehat{a}}=0$.

In this paper, we consider the two important problems. One aim is to generalize
the methods of the Foldy-Wouthuysen transformations, that we previously used for
the analysis of the spin in static gravitational fields, to the case of the
stationary gravitational configurations. Another aim is to systematically
investigate the dependence of the spin dynamics on the choice of a tetrad.
In particular, we derive the general result for the angular velocity
of spin precession that is valid for an arbitrary tetrad gauge.

The paper is organized as follows. In Sec.~\ref{Hamiltonian} we consider the
Dirac equation in a weak gravitational field of a rotating source. The Hermitian
Hamiltonian is derived. Sec.~\ref{FW} presents the derivation of the precession
angular velocity of spin in a stationary gravitational field. The dynamics of
a classical spin is analyzed in Sec.~\ref{CS} where we find a general expression
for the precession velocity of the physical spin in arbitrary external classical
fields. These results are then applied in Sec.~\ref{EP} to the derivation of the
classical spin dynamics in arbitrary gravitational field configurations.
Sec.~\ref{tetrad} is devoted to the analysis of the dependence of the spin
precession effect on the choice of a tetrad. We find a general equation that
makes it possible to directly compare results obtained in the literature
in different tetrad gauges. Specifically, we demonstrate the complete agreement
of the classical and quantum spin dynamics in the Schwinger gauge. In Sec.~\ref{rot}
we show that our general results can be also used for the study of the motion of
spin in the flat spacetime for the rotating reference frame. Our derivations
confirm the earlier observations obtained on the basis of the Thomas precession
arguments. Finally, in Sec.~\ref{discussion} we draw the conclusions.

We denote world indices by Latin letters $i,j,k,\dots = 0,1,2,3$
and reserve first Greek letters for tetrad indices, $\alpha,\beta,\dots =
0,1,2,3$. Spatial indices are denoted by Latin letters from the
beginning of the alphabet, $a,b,c,\dots = 1,2,3$.
The separate tetrad indices are distinguished by hats.

%%%%%%%%%%%%%%%%%%%%%%%%%%%%%%%%%%%%%%%%%%%%%%%%%
\section{Dirac Hamiltonian for a stationary metric}\label{Hamiltonian}
%%%%%%%%%%%%%%%%%%%%%%%%%%%%%%%%%%%%%%%%%%%%%%%%%

The approximate gravitational field of a rotating body at a large distance
is described by the Lense-Thirring (LT) metric \cite{LT}:
\begin{eqnarray}
ds^2 &=& \left[ 1 - {\frac {2GM} {c^2\rho}} \right]c^2dt^2
-\,a\sin^2\!\theta\, {\frac {4GM} {c^2\rho}}
cdt\,d\phi\nonumber\\
&& -\left[1 + {\frac {2GM} {c^2\rho}} \right]d\rho^2
-\,\rho^2\left[d\theta^2 + \sin^2\theta d\phi^2\right].
\label{asymKerr}
\end{eqnarray}
With the help of the coordinate transformation
\begin{equation}
\rho = r\left(1 + {\frac {GM} {2c^2\,r}}\right)^2,
\end{equation}
one can bring the line element to the isotropic form and
subsequently use the Cartesian coordinate system. The final form
of the line element is given by
\begin{equation}\label{LT}
ds^2 = V^2c^2dt^2 - W^2\,\delta_{ab}\,(dx^a - K^acdt)\,(dx^b - K^bcdt),
\end{equation}
with
\begin{eqnarray}
V &=& \left(1 - {\frac {GM}{2c^2r}}\right)
\left(1 + {\frac {GM} {2c^2r}}\right)^{-1},\label{Vk}\\
W &=& \left(1 + {\frac {GM} {2c^2r}}\right)^2,\label{Wk}\\
K^a &=& {\frac 1c}\,\epsilon^{abc}\,\omega_b\,x_c.\label{Ua}
\end{eqnarray}
The non-diagonal components of the metric (that reflect the rotation
of the source) are described by the so-called Kerr field $\bm K$
that is given by Eq. (\ref{Ua}) with
\begin{equation}
\bm\omega =\frac{2G}{c^2r^3}\bm J= \left(0,\quad 0,\quad
\frac{2GM\,a}{c\,r^3}\right),\label{omega}
\end{equation}
where $\bm J=Mca\bm e_z$ is the total angular momentum of the source.

The {\it exact} metric of the flat spacetime seen by an
accelerating and rotating observer also has form (\ref{LT}).
In the latter case \cite{HN}, however,
\begin{equation}
V = 1 + {\frac {{\bm a}\cdot{\bm r}}{c^2}},\qquad W = 1,\qquad
K^a =-{\frac 1c}\,(\bm\omega\times\bm r)^a,\label{VWni}
\end{equation}
where ${\bm a}$ describes acceleration of the observer and $\bm\omega$
is an angular velocity of a noninertial reference system. Both are
independent of the spatial coordinates, but may depend arbitrarily
on time $t$.

The similarity between the two cases is not occasional. Lense and Thirring
have discovered in 1918 that rotating bodies ``drag" the spacetime around
themselves (frame dragging \cite{LT}). In other words, they have demonstrated
the equivalence between rotating frames and spacetimes created by rotating
bodies. In the weak-field approximation, the motion of particles in a
gravitational field of a rotating source is identical to their motion
in a noninertial frame rotating with the angular
velocity (see, e.g., Ref. \cite{LL}) % p. 328
\begin{equation}
\bm\omega = \frac {c}{2}\,{\rm curl}\,\bm g,\qquad g_a=-g_{0a}.\label{equiv}
\end{equation}

Let us choose the orthonormal tetrad
\begin{equation}
e_i^{\,\widehat{0}} = V\,\delta^{\,0}_i,\qquad
e_i^{\widehat{a}} =
W\left( \delta^a_i - K^a\,\delta^{\,0}_i\right),\qquad a,b =1,2,3.
\label{coframe}
\end{equation}

The covariant Dirac equation for spin-1/2 particles in curved
spacetimes has the form
\begin{equation}
(i\hbar\gamma^\alpha D_\alpha - mc)\psi=0,\qquad \alpha=0,1,2,3.
\label{Dirac0}
\end{equation}
The Dirac matrices $\gamma^\alpha$ are defined in
local Lorentz (tetrad) frames. The spinor
covariant derivatives are given by
\begin{equation}
D_\alpha = e_\alpha^i D_i,\qquad D_i = \partial _i + {\frac
i4}\sigma_{\alpha\beta}\Gamma_i{}^{\alpha\beta},\label{eqin2}
\end{equation}
where $\Gamma_i{}^{\alpha\beta} = - \Gamma_i{}^{\beta\alpha}$ are the
Lorentz connection coefficients,
$\sigma^{\alpha\beta} = i(\gamma^\alpha\gamma^\beta -\gamma^\beta
\gamma^\alpha)/2$. Eqs. (\ref{Dirac0}),(\ref{eqin2}) show that the
gravitational and inertial effects are encoded in coframes
(see Refs. \cite{Ob1,Ob2} and references therein).

Eq. (\ref{Dirac0}) is recast into the familiar Schr\"odinger form
\begin{equation}\label{Dirac1}
i\hbar{\frac {\partial\psi} {\partial t}} = %\widehat
{\cal H}\,\psi
\end{equation}
with the Hamilton operator
\begin{eqnarray}
{\cal H} &=& \beta mc^2V + {\frac V W}\,c(\bm{\alpha} \cdot\bm{p})
- {\frac {i\hbar c}{2W}}\left(\bm{\alpha}\cdot \bm{\nabla}V\right)
- {\frac {i\hbar cV}{W^2}}\left(\bm{\alpha}
\cdot\bm{\nabla}W\right)\nonumber\\
&& -\,i\hbar c \bm{K}\cdot\bm{\nabla} - {\frac {i\hbar c}
2}\,(\bm{\nabla} \cdot\bm{K}) - {\frac {3i\hbar c}
{2W}}\,(\bm{K}\cdot\bm{\nabla}W) + {\frac {\hbar
c}4}\,(\bm{\nabla}\times\bm{K})\cdot\bm{\Sigma}.\label{Hamilton0}
\end{eqnarray}
Here $\bm p=-i\hbar\nabla$, and we remind that $\beta = \gamma^{\hat 0},
{\bm\alpha} = \{\alpha^a\}, {\bm\Sigma} = \{\Sigma^a\}$, where the
3-vector-valued Dirac matrices have their usual form: $\alpha^a
= \gamma^{\hat 0}\gamma^a$ and $\Sigma_a = i\epsilon_{abc}\gamma^b
\gamma^c/2$ ($a,b,c,\dots = 1,2,3$). Redefining the spinor field and the
Hamiltonian,
\begin{equation}
\psi' = W^{3/2}\,\psi, %\qquad \widehat{\cal H}' = W^{3/2}
%\,\widehat{\cal H}\,W^{-\,{3/2}},\label{psi1}
\end{equation}
we obtain the new Hamiltonian (which is explicitly Hermitian with
respect to the usual flat space scalar product):
\begin{eqnarray}
%\widehat
{\cal H}' &=& \beta mc^2V + {\frac c 2}\left[(\bm{\alpha}
\cdot\bm{p}){\cal F} + {\cal F}(\bm{\alpha}\cdot\bm{p})\right]\nonumber\\
&& +\,{\frac c2}\left(\bm{K}\cdot\bm{p} + \bm{p}\cdot\bm{K}\right)
+ {\frac {\hbar
c}4}\,(\bm{\nabla}\times\bm{K})\cdot\bm{\Sigma}.\label{Hamilton1}
\end{eqnarray}
Here ${\cal F} = V/W$.

Substituting (\ref{Vk})-(\ref{omega}) into (\ref{Hamilton1}), we
find:
\begin{eqnarray}
{\cal H}' &=& \beta mc^2V + {\frac c 2}\left[(\bm{\alpha}\cdot\bm{p})
{\cal F} + {\cal F}(\bm{\alpha}\cdot\bm{p})\right]\nonumber\\
&& +\frac{2G}{c^2r^3}\bm{J}\cdot\left(\bm{r}\times\bm{p}\right)
+ \frac{\hbar G}{2c^2r^3}\left[ {\frac {3(\bm{r}\cdot\bm{J})(\bm{r}
\cdot\bm{\Sigma})}{r^2}} -\bm{J}\cdot\bm{\Sigma}\right].\label{Hamilton2}
\end{eqnarray}
Note that the angular momentum operator $\bm{l}=\bm{r}\times\bm{p}$
commutes with $\bm{\omega}$ which depends on the radius. Dirac
Hamiltonian (\ref{Hamilton2}) contains the first part describing the
static gravitational field and the second one characterizing the
contribution of rotation of the central body.

%%%%%%%%%%%%%%%%%%%%%%%%%%%%%%%%%%%%%%%%%%%%%%%%%
\section{Foldy-Wouthuysen Hamiltonian and operator equations of motion}
\label{FW}
%%%%%%%%%%%%%%%%%%%%%%%%%%%%%%%%%%%%%%%%%%%%%%%%%

To obtain the FW Hamiltonian, we perform the FW transformation by
the method developed in Refs. \cite{JMP,PRA}. In the weak field
approximation, there are three small parameters:
\begin{equation}
|V-1|\ll 1,\qquad |{\cal F}-1|\ll 1,\qquad |\bm K|\ll 1.
\end{equation}
Evidently, any bilinear combinations of these parameters can be
neglected. The FW Hamiltonian can be presented as a sum of a free
particle Hamiltonian and terms proportional to $|V-1|,~|{\cal
F}-1|$, and $\bm K$. Only the last term, ${\cal H}_{FW}^{(2)}$,
defines the contribution of rotation of the central body, while
the other terms characterize the FW Hamiltonian of the particle in
a static gravitational field. The rotation-independent contribution
${\cal H}_{FW}^{(1)}$ was calculated earlier \cite{PRD}:
\begin{widetext}
\begin{eqnarray}
{\cal H}_{FW}^{(1)}=\beta\epsilon + \frac{\beta}{2}\left\{
\frac{m^2c^4}{\epsilon },V-1\right\} + \frac{\beta}{2}\left\{
\frac{c^2\bm p^2}{\epsilon },{\cal F}-1\right\}-\frac{\beta
\hbar mc^4}{4 \epsilon (\epsilon + mc^2)}\biggl[\bm{\Sigma}
\cdot(\bm\phi\times\bm p)\nonumber\\
- \bm{\Sigma}\cdot(\bm p\times\bm\phi)+ \hbar\nabla\!
\cdot\!\bm\phi\biggr] + \frac{\beta \hbar^2 mc^6(2\epsilon^3
+ 2\epsilon ^2mc^2 + 2\epsilon m^2c^4+m^3c^6)}{8\epsilon^5
(\epsilon +mc^2)^2}(\bm p\cdot\!\nabla)(\bm p\cdot\!\bm\phi)
\nonumber\\
+ \frac{\beta\hbar c^2}{4\epsilon }\left[\bm{\Sigma}\cdot
(\bm f\times\bm p) - \bm{\Sigma}\cdot(\bm p\times\bm f)+\hbar
\nabla\! \cdot\!\bm f\right]-\frac{\beta \hbar^2 c^4(\epsilon^2
+m^2c^4)}{4\epsilon^5}(\bm p\cdot\!\nabla)(\bm p\cdot\!\bm f).
\label{eq7}
\end{eqnarray}
\end{widetext}
Here $\epsilon=\sqrt{m^2c^4+c^2\bm p^2}$ and the curly bracket
$\{\cdots,\cdots\}$ denotes the anticommutator. We also use the
notation of \cite{Ob1,PRD} for the gradients: ${\bm \phi} =
\{\partial_a V\}, {\bm f} = \{\partial_a {\cal F}\}$, $a = 1,2,3$.

To find the rotation-dependent term ${\cal H}_{FW}^{(2)}$, it is
sufficient to keep the leading term in the FW transformation
operator \cite{PRD} corresponding to the free particle
transformation:
\begin{equation}
U=\frac{\epsilon+mc^2+\beta c\bm{\alpha}
\cdot\bm{p}}{\sqrt{2\epsilon(\epsilon+mc^2)}}. \label{fpFWt}
\end{equation}
Corrections to this approximation can be neglected because they
only affect terms in the FW Hamiltonian which are bilinear in
small parameters $|V-1|,~|{\cal F}-1|$, and $\bm K$.

This FW transformation leads after straightforward but tedious
calculations to the final FW Hamiltonian which is given by
\begin{equation}
\begin{array}{c}
{\cal H}_{FW}={\cal H}_{FW}^{(1)}+{\cal H}_{FW}^{(2)},\quad
{\cal H}_{FW}^{(2)}=\frac{2G}{c^2r^3}\bm J\cdot\bm l + {\frac
{\hbar G}{2c^2r^3}}\left[\frac{3(\bm r\cdot\bm J)(\bm r\cdot
\bm \Sigma)}{r^2}- \bm J\cdot\bm \Sigma\right]\\
-\frac{3\hbar G}{8}\left \{\frac{1}{\epsilon(\epsilon+mc^2)},
\left[\frac{2\{(\bm J\cdot\bm l), (\bm \Sigma\cdot\bm l)\}}{r^5}
+\frac{1}{2}\left\{\left(\bm \Sigma\cdot (\bm p\times\bm l)
-\bm\Sigma\cdot(\bm l\times\bm p)\right),\frac {(\bm r\cdot\bm J)}
{r^5}\right\}\right.\right.\\
\left.\left. +\left\{\bm \Sigma\cdot(\bm p\times(\bm p\times\bm
J)), \frac{1}{r^3}\right\}\right]\right\}-\frac{3\hbar^2c^2
G}{8}\left\{ (5p_r^2-\bm p^2)\frac{2\epsilon^2+\epsilon
mc^2+m^2c^4}{\epsilon^4 (\epsilon+mc^2)^2},\frac{(\bm J\cdot\bm
l)}{r^5}\right\},
\end{array} \label{finalH}
\end{equation}
where $\bm l=\bm r\times\bm p$ is an angular momentum operator,
and the operator $p_r^2=-\frac{\hbar^2}{r^2}\frac{\partial}{\partial
r}\left(r^2\frac{\partial}{\partial r}\right)$ is proportional to
the radial part of the Laplace operator. The equation of rotation
of the spin is obtained via commuting the FW Hamiltonian with the
polarization operator $\bm\Pi=\beta\bm\Sigma$ and is given by
\begin{equation}
\frac{d\bm \Pi}{dt}=\frac{i}{\hbar}[{\cal H}_{FW},\bm
\Pi]=\bm\Omega^{(1)}\times\bm \Sigma+\bm\Omega^{(2)}\times\bm \Pi,
\label{spinmeq}\end{equation} where $\bm\Omega^{(1)}$ is the
operator of angular velocity of rotation of the spin in the static
gravitational field derived in Ref. \cite{PRD},
\begin{equation}
\bm\Omega^{(1)}=-\frac{mc^4}{\epsilon (\epsilon
+mc^2)}\left(\bm\phi\times\bm
p\right)+\frac{c^2}{\epsilon}\left(\bm f\times\bm
p\right),\label{eqol}
\end{equation}
and the newly obtained contribution from the LT effect is equal to
\begin{eqnarray}
\bm\Omega^{(2)} &=& \frac{G}{c^2r^3} \left[\frac{3(\bm r\cdot\bm
J)\bm r} {r^2} - \bm J\right]
-\frac{3G}{4}\left\{\frac{1}{\epsilon (\epsilon +
mc^2)},\left[\frac{2\{\bm l,(\bm J\cdot\bm l)\}}{r^5}
\right.\right.\nonumber\\
&& \left.\left. \,+\,\frac{1}{2}\left\{(\bm p\times\bm l - \bm l
\times\bm p), \frac{(\bm r\cdot\bm J)}{r^5}\right\} + \left\{(\bm
p \times(\bm p \times\bm J)),\frac{1}{r^3}\right\}\right]\right\}.
\label{finalOmega}
\end{eqnarray}

The second term on the right-hand side of Eq. (\ref{spinmeq}) contains
an additional $\beta$ factor as compared to the first term. This
is a manifestation of the gravitoelectric and the gravitomagnetic
origin of the static gravitational field and of the Kerr (Lense-Thirring)
field, respectively. The equation of spin motion in the electromagnetic
field has a similar form (see Eq. (36) in Ref. \cite{PRD}). The difference
between the two terms on the right-hand side of Eq. (\ref{spinmeq}) is
caused by the fact that $\bm\Omega^{(1)}$ should contain the velocity
operator rather than the momentum one. Since the velocity operator is
proportional to an additional $\beta$ factor and is equal to $\bm v
= \beta c\bm p/\epsilon$ for free particles, the operator $\bm\Omega^{(1)}$,
expressed in terms of $\bm v$, also acquires an additional $\beta$
factor.

In Eq. (\ref{finalH}), the Hamiltonian ${\cal H}_{FW}$ defines the
zero component of the covariant four-momentum operator, while its
spatial components are expressed by the operator $\bm p$ taken
with the opposite sign:
$$p_i=i\hbar\frac{\partial}{\partial x^i}=
\left(\frac{{\cal H}_{FW}}{c},-\bm p\right).$$

The equation of motion of the particle defines the evolution of
the contravariant four-momentum operator which spatial components
($a,b=1,2,3$) are given by
$$p^a = g^{ab}p_b + g^{0a}p_0.$$

In a stationary metric, the evolution of the contravariant
momentum operator in the weak field approximation is defined by
\begin{eqnarray}
F^a=\frac{dp^a}{dt}=-\frac{dp_a}{dt}+\frac14\left\{\left\{v^b,
\frac{\partial g^{ai}}{\partial x^b}\right\},p_i\right\},\qquad
\frac{d\bm p}{dt}=\frac{i}{\hbar}[{\cal H}_{FW},\bm p],
\label{eqFi}\end{eqnarray} where $F^a$ is the force operator and
$v^a\approx\beta c^2p^a/\epsilon \approx c^2p^a/{\cal H}_{FW}$ is
the velocity operator.

One can calculate the force operator caused by the LT effect
without allowance for contributions from $V,W$. This operator is
equal to
\begin{eqnarray}
\bm F=\frac c2 \left({\rm curl}\,\bm K\times\bm p-\bm p\times
{\rm curl}\,\bm K\right)+\bm F_s,\label{finalpi}
\end{eqnarray}
where
\begin{eqnarray}
{\rm curl}\,\bm K=\frac{2G}{c^3r^3} \left[\frac{3(\bm r\cdot
\bm J)\bm r} {r^2} - \bm J\right],\quad %\nonumber\\
\bm F_s=-\nabla\left(\frac{\hbar G}{2c^2r^3} \left[\frac{3(\bm
r\cdot\bm J)(\bm r\cdot\bm \Sigma)}{r^2}- \bm J\cdot\bm
\Sigma\right]\right.\nonumber\\-\frac{3\hbar G }{8}\left\{\frac
{1}{\epsilon(\epsilon+mc^2)},\left[\frac{2\{(\bm J\cdot\bm l),
(\bm \Sigma\cdot\bm l)\}}{r^5}+\frac{1}{2} \left\{\left(\bm\Sigma
\cdot (\bm p\times\bm l)-\bm \Sigma\cdot(\bm l\times\bm p)\right),
\frac {(\bm r\cdot\bm J)}{r^5}\right\}\right.\right.\nonumber\\
\left.\left.\left. +\left\{\bm \Sigma\cdot(\bm p\times(\bm
p\times\bm J)), \frac{1}{r^3}\right\}\right]\right\}\right).
\label{finlK}
\end{eqnarray}
The operator equation (\ref{finalpi}) for the small spin-dependent
force $\bm F_s$ is in the best compliance with the corresponding
classical equation \cite{LL}. Since the Dirac spin operator is
$\bm s=\hbar\bm \Sigma/2$, the Eqs. (\ref{finalpi}), (\ref{finlK})
yield the corresponding semiclassical equation:
\begin{eqnarray}
\bm{\mathcal{F}}=c\,{\rm curl}\,\bm K\times\bm
p+\bm{\mathcal{F}}_s,\label{finalpt}
\end{eqnarray}
\begin{eqnarray}
\bm{\mathcal{F}}_s=-\nabla\left(\frac{G}{c^2r^3} \left[\frac{3(\bm
r\cdot\bm J)(\bm r\cdot\bm s)}{r^2}- \bm J\cdot\bm s\right]\right.\nonumber\\
\left.-\frac{3G}{\epsilon(\epsilon+mc^2)}\left[\frac{2(\bm J\cdot\bm l)(\bm
s\cdot\bm l)}{r^5}+\frac{\left(\bm s \cdot [\bm p\times\bm l]\right)
(\bm r\cdot\bm J)}{r^5} + \frac{\left(\bm s \cdot [\bm p\times[\bm
p\times\bm J]]\right)}{r^3}\right]\right). \label{finlt}
\end{eqnarray}
Our relativistic result (\ref{finlK}), (\ref{finlt}) for the
spin-dependent force perfectly agrees with the corresponding
nonrelativistic classical formulas previously obtained in Ref.
\cite{wald} on the basis of the Mathisson-Papapetrou equations
\cite{Mathisson,Papapetrou}.

Our quantum Eqs. (\ref{finalpi})-(\ref{finlt}) actually agree with the
classical results of both Mathisson-Papapetrou and Pomeransky-Khriplovich
approaches. This follows from the fact that the spin-dependent
part of Hamiltonian has the form ${\cal H}_s=\hbar(\bm\Omega^{(1)}
\cdot\bm \Sigma+\bm\Omega^{(2)}\cdot\bm\Pi)/2$ that perfectly agrees
with the general classical Eq. (47) of the Ref. \cite{PK}.
This is also in accordance with the earlier attempts (see Ref.
\cite{wong}, for example) to establish a direct general
correspondence between the quantum dynamics and the classical
equations of motion of the Mathisson-Papapetrou type.

The semiclassical formula corresponding to Eq. (\ref{finalOmega})
and describing the motion of average spin has the form
\begin{equation}\label{OmegaVt}
\bm\Omega^{(2)}= \frac{G}{c^2r^3} \left[\frac{3(\bm r\cdot\bm J)\bm r}
{r^2} - \bm J\right] - \frac{3G}{r^3\epsilon(\epsilon+mc^2)}\left[
\frac{2\bm l (\bm J\cdot\bm l)+(\bm p\times\bm l) (\bm r\cdot\bm J)}
{r^2}+\bm p\times(\bm p \times\bm J)\right].
\end{equation}
In a nonrelativistic approximation, the Eq. (\ref{OmegaVt}) coincides
with the equation obtained by Schiff \cite{Schiff}. The second
term in the Eq. (\ref{OmegaVt}) describes relativistic corrections.
The Eq. (\ref{OmegaVt}) can also be expressed in the equivalent form:
\begin{equation}\label{OmegaCl}
\bm\Omega^{(2)}=\frac{G}{c^2r^3}\left[\frac{3(\bm r\cdot\bm J)\bm
r}{r^2} - \bm J\right] -
\frac{3G}{r^5\epsilon(\epsilon+mc^2)}\left[\bm l (\bm l \cdot\bm
J)+(\bm r\cdot\bm p) (\bm p\times(\bm r\times\bm J))\right].
\end{equation}

The quantum mechanical and semiclassical equations (\ref{finalH}),
(\ref{finalOmega}), (\ref{finalpi})-(\ref{OmegaCl})
are principal new results.

%%%%%%%%%%%%%%%%%%%%%%%%%%%%%%%%%%%%%%%%%%%%%%%%%
\section{Classical spin in external fields}\label{CS}
%%%%%%%%%%%%%%%%%%%%%%%%%%%%%%%%%%%%%%%%%%%%%%%%%

The dynamical equations that determine the motion of a spinning particle
in external classical fields can be written, quite generally, in the form
\begin{eqnarray}
{\frac {dU^\alpha}{d\tau}} &=& {\cal F}^\alpha,\label{dotU}\\
{\frac {dS^\alpha}{d\tau}} &=& \Phi^\alpha{}_\beta S^\beta.\label{dotS}
\end{eqnarray}
The forces ${\cal F}^\alpha$ are determined by the external fields
(electromagnetic, gravitational, etc.) acting on a particle. Similarly,
the spin is affected by the external fields through a spin transport
matrix $\Phi^\alpha{}_\beta$. Normalization of the velocity, $U_{\alpha}
U^\alpha = c^2$, and its orthogonality to the spin, $S_{\alpha}U^\alpha = 0$,
impose on the right-hand sides of (\ref{dotU}),(\ref{dotS}) the conditions
\begin{equation}
U_\alpha {\cal F}^\alpha = 0,\qquad U_\alpha\Phi^\alpha{}_\beta S^\beta =
S_\alpha {\cal F}^\alpha.\label{cc}
\end{equation}
Since
\begin{equation}
S_\alpha\Phi^\alpha{}_\beta S^\beta={\frac 12}\,{\frac{d(S_\alpha S^\alpha)}
{d\tau}}=0,
\end{equation}
the spin transport matrix is skew-symmetric: $\Phi_{\alpha\beta}=-\Phi_{\beta\alpha}$.

The components of the 4-velocity are conveniently parametrized by the spatial
3-velocity $v^a$ ($a = 1,2,3$) as
\begin{equation}\label{U}
U^\alpha = \left(\begin{array}{c}\gamma \\ \gamma v^a\end{array}\right),
\end{equation}
where $\gamma = (1 - v^2/c^2)^{-1/2}$ is the Lorentz factor ($v^2 = \delta_{ab}
v^av^b$). When the particle is at rest, $v^a =0$, its 4-velocity reduces to
\begin{equation}\label{Ur}
u^\alpha = \delta^\alpha_0 = \left(\begin{array}{c}1 \\ 0\end{array}\right).
\end{equation}
The actual 4-velocity $U^\alpha$ is obtained from the rest-frame components
with the help of the Lorentz transformation $U^\alpha = \Lambda^\alpha{}_\beta
u^\beta$ where
\begin{equation}\label{Lambda}
\Lambda^\alpha{}_\beta = \left(\begin{array}{c|c}\gamma & \gamma v_b/c^2 \\
\hline \gamma v^a & \delta^a_b + (\gamma - 1)v^av_b/v^2\end{array}\right).
\end{equation}
Hereafter the Latin indices from the beginning of the alphabet ($a,b,\dots
= 1,2,3$ which label the spatial components of the objects) are raised and
lowered with the help of the Euclidean 3-dimensional metric $\delta_{ab}$.

The physical components of spin $s^\alpha$ are defined in the {\it rest frame}
of a particle. Accordingly, we have $S^\alpha = \Lambda^\alpha{}_\beta s^\beta$.
The dynamical equation for the {\it physical spin} is derived by substituting
this relation into (\ref{dotS}) which yields
\begin{equation}
{\frac {ds^\alpha}{d\tau}} = \Omega^\alpha{}_\beta s^\beta.\label{ds}
\end{equation}
Here we introduced $\Omega^\alpha{}_\beta = \phi^\alpha{}_\beta +
\pi^\alpha{}_\beta$ where
\begin{equation}
\phi^\alpha{}_\beta = (\Lambda^{-1})^\alpha
{}_\gamma\Phi^\gamma{}_\delta\Lambda^\delta{}_\beta,\qquad
\pi^\alpha{}_\beta = -\,(\Lambda^{-1})^\alpha{}_\gamma{\frac d {d\tau}}
\Lambda^\gamma{}_\beta.\label{Thomas}
\end{equation}
The physical spin has only three spatial components. One can verify that
the 0-th component of (\ref{ds}) is identically satisfied [in fact, it is
identical to the second compatibility condition (\ref{cc})]. As a result,
the dynamical equation (\ref{ds}) reduces to the 3-vector form
\begin{equation}
{\frac {ds^a}{d\tau}} = \Omega^a{}_b s^b,\qquad {\rm or}\qquad
{\frac {d{\bm s}}{d\tau}} = {\bm \Omega}\times{\bm s}.\label{ds1}
\end{equation}
Here ${\bm \Omega} = {\bm \phi} + {\bm \pi}$ and the components
of the 3-vectors are introduced by ${\bm s} = (s^1, s^2, s^3)$
and ${\bm \Omega} = (\Omega_{32}, \Omega_{13}, \Omega_{21})$.

The new general equations (\ref{Thomas})--(\ref{ds1}) are valid for
a spinning particle interacting with any external fields. In the next
section, we specify these equations to the case of the gravitational field.

%%%%%%%%%%%%%%%%%%%%%%%%%%%%%%%%%%%%%%%%%%%%%%%%%
\section{Equivalence principle and dynamics of classical and quantum spins
in curved spacetimes}\label{EP}
%%%%%%%%%%%%%%%%%%%%%%%%%%%%%%%%%%%%%%%%%%%%%%%%%

The equivalence principle (EP) is known to be the cornerstone of
general relativity. The EP results in the general equation of motion
of classical test particles in curved spacetimes:
\begin{equation}
{\frac{DU^\alpha}{d\tau}} = 0,\label{MPA1}
\end{equation}
where $D/(d\tau)$ denotes the covariant derivative along the curve.
The corresponding equation of motion of the four-component spin used
in Refs.~\cite{PK,Dvornikov} and many other works is very similar:
\begin{equation}
{\frac{DS^\alpha}{d\tau}}=0.\label{eqPK}
\end{equation}

In the present work, we do not consider a relatively weak influence
of the spin on particle's trajectory produced by the
Mathisson-Papapetrou force \cite{PK,Mathisson,Papapetrou} which
results in a weak violation of the equivalence principle by
the curvature-dependent terms \cite{Plyatsko}. For the Kerr
spacetime, the deviation from the geodetic motion under the
influence of spin was recently comprehensively studied in the
framework of the analytic perturbation approach in \cite{singh},
see also the relevant references there.

In the context of our present investigation of the {\it dynamics of spin},
it is worthwhile to stress that the account of the Mathisson-Papapetrou
terms does not change the spin dynamics in the current approximation.
There is thus no any difference between the Mathisson-Papapetrou and
Pomeransky-Khriplovich approaches within our framework. Nevertheless
we find it more convenient to refer specifically to \cite{PK} where
the exact equation of motion for the three-component spin was
obtained in explicit form.

The Eq. (\ref{MPA1}) states identical motion of all classical
particles in curved spacetimes. Similarly, the Eq. (\ref{eqPK})
states identical motion of all classical spins (gyroscopes). This
important conclusion leads to a great difference between dynamics
of the spin in electrodynamics and gravity. Angular velocities of
precession of all classical and quantum spins moving with the same
velocity in the curved spacetime are equal. Thus, spinning
particles cannot have any anomalous gravitomagnetic moments
\cite{KO}. It has been proved that both the anomalous
gravitomagnetic moment and the gravitoelectric dipole moment being
gravitational analogs of the anomalous magnetic moment and the
electric dipole moment, respectively, are identically zero
\cite{KO}. Relations obtained by Kobzarev and Okun predict equal
frequencies of precession of classical and quantum spins in any
curved spacetimes \cite{T1,PRD2}. Nevertheless, this conclusion
was discussed for a long time (see Refs.
\cite{Mashhoon2,PRD,PRD2,Warszawa} and references therein).

On the contrary, angular velocities of spin precession of different
particles which are determined by the Thomas-Bargmann-Michel-Telegdi
equation \cite{Thoms,BMT} do not coincide and, generally speaking,
differ from an angular velocity of precession of a classical rotator.

Comparing (\ref{MPA1})-(\ref{eqPK}) with (\ref{dotU}) and (\ref{dotS}),
we find the explicit expressions for the force and the spin transport
matrix
\begin{equation}
{\cal F}^\alpha = \Phi^\alpha{}_\beta U^\beta,\qquad \Phi^\alpha{}_\beta
= -\,U^i\Gamma_{i\beta}{}^\alpha,\label{forceG}
\end{equation}
in terms of the gravitational field $\Gamma_{i\beta}{}^\alpha$.
Using this in (\ref{Thomas}), we find explicitly
\begin{eqnarray}
\phi^{ab} &=& -\,U^i\left[\Gamma_i{}^{ba} + {\frac {\gamma^2}{\gamma + 1}}
\,{\frac {v_c}{c^2}}\left(\Gamma_{ic}{}^av^b - \Gamma_{ic}{}^bv^a\right) +
{\frac {\gamma}{c^2}}\left(\Gamma_{i\widehat{0}}{}^av^b
- \Gamma_{i\widehat{0}}{}^b v^a\right)\right],\\
\pi^{ab} &=& U^i\,{\frac {\gamma^2}{\gamma + 1}}\left[{\frac {v_c}{c^2}}
\left(\Gamma_{ic}{}^av^b - \Gamma_{ic}{}^bv^a\right) + {\frac 1 {c^2}}\left(
\Gamma_{i\widehat{0}}{}^av^b - \Gamma_{i\widehat{0}}{}^bv^a\right)\right].
\end{eqnarray}
Hence the precession of the physical spin in the gravitational field is
described by
\begin{equation}
\Omega_{a} = \epsilon_{abc}\,U^i\left({\frac 12}\Gamma_i{}^{cb} + {\frac
{\gamma}{\gamma + 1}}\,\Gamma_{i\widehat{0}}{}^bv^c/c^2\right).\label{OmegaG}
\end{equation}
This exact formula can be used also in the flat spacetime for noninertial
reference frames, since the connection $\Gamma_{i\beta}{}^\alpha$ contains
information about both gravitational and inertial effects.

The Eq. (\ref{OmegaG}) has been first obtained by Pomeransky and Khriplovich
\cite{PK} as a result of a comparison of the equations of motion of spin in
electrodynamics and gravity, and more recently has been consistently derived
by Dvornikov \cite{Dvornikov}. Note that unlike Ref. \cite{Dvornikov},
our results can be easily extended to any external fields (electromagnetic,
gravitational, scalar, and other).

%%%%%%%%%%%%%%%%%%%%%%%%%%%%%%%%%%%%%%%%%%%%%%%%%
\section{Classical spin in nonstatic spacetimes}\label{tetrad}
%%%%%%%%%%%%%%%%%%%%%%%%%%%%%%%%%%%%%%%%%%%%%%%%%

Description of a spin requires the introduction of a tetrad (the frame
$e_\alpha = e_\alpha^i\partial_i$ and the dual coframe $\vartheta^\alpha
= e^\alpha_idx^i$). In physical terms a choice of a tetrad means a
selection of a local reference system of an observer.

Mathematically, there are infinitely many tetrads since a reference frame
of an observer can obviously be constructed in an infinitely many ways.
In particular, {}from a given tetrad field $e{}^\alpha_i$ we can obtain
a continuous family of tetrads by performing the Lorentz transformation
$e'{}^\alpha_i = \Lambda^\alpha{}_\beta e^\beta_i$, where the elements of
the Lorentz matrix $\Lambda^\alpha{}_\beta(x)$ are arbitrary functions of the
spacetime coordinates. In practice, there are three most widely used gauges.

{\it Schwinger gauge}. Probably for the first time introduced independently
by Schwinger \cite{Schwinger} and Dirac \cite{dirac} (and widely used in many
works, including \cite{HN} and our current study), this choice demands that
the tetrad matrix $e^\alpha_i$, and its inverse $e^i_\alpha$, both have the
trivial elements in the upper-right blocks:
\begin{equation}\label{Sgauge}
e^\alpha_i = \left(\begin{array}{c|c} e^{\widehat 0}_0 & 0 \\
\hline e^{\widehat a}_0 & e^{\widehat a}_b\end{array}\right),\qquad
e_\alpha^i = \left(\begin{array}{c|c} e_{\widehat 0}^0 & 0 \\
\hline e^a_{\widehat 0} & e_{\widehat b}^a\end{array}\right).
\end{equation}

{\it Landau-Lifshitz gauge} (see, e.g., Ref. \cite{LLp}) fixes the
tetrad matrices so that they both have the trivial elements in the
lower-left blocks:
\begin{equation}\label{Lgauge}
e^\alpha_i = \left(\begin{array}{c|c} e^{\widehat 0}_0 & e^{\widehat 0}_b \\
\hline 0 & e^{\widehat a}_b\end{array}\right),\qquad
e_\alpha^i = \left(\begin{array}{c|c} e_{\widehat 0}^0 & e^0_{\widehat b} \\
\hline 0 & e_{\widehat b}^a\end{array}\right).
\end{equation}

{\it Symmetric gauge}. Using the Minkowski flat metric $g_{\alpha\beta} =
{\rm diag}(c^2, -1, -1, -1)$, we can move the anholonomic index down and
construct the matrix $e_{\alpha i} := g_{\alpha\beta}e^\beta_i$. The tetrad
is called symmetric  (hence the name, symmetric gauge) when the resulting
matrix is invariant under the transposition operation which we symbolically
can write as
\begin{equation}
e_{\alpha i} = e_{i\alpha}.\label{Kgauge}
\end{equation}
Such a tetrad was used by Pomeransky and Khriplovich \cite{PK} and Dvornikov
\cite{Dvornikov}.

In the framework of our current study, we choose the {\it Schwinger gauge}
by specifying the coframe as (\ref{coframe}).

The other tetrads are obtained from our $e^\alpha_i$ with
the help of the Lorentz transformation $e'{}^\alpha_i = \Lambda^\alpha{}_\beta
e^\beta_i$, where
\begin{equation}\label{Lorentz}
\Lambda^\alpha{}_\beta = \left(\begin{array}{c|c}\lambda & \lambda q_b/c \\
\hline \lambda cq^a & \delta^a_b + (\lambda - 1)q^aq_b/q^2\end{array}\right).
\end{equation}
Here we denote
\begin{equation}\label{qxi}
q^a = \xi\,{\frac {WK^a}{V}},\qquad \lambda = {\frac {1}{\sqrt{1 - q^2}}}.
\end{equation}
(As usual, $q^2 = \delta_{ab}q^aq^b$). The constant $\xi$ conveniently
parametrizes different choices of tetrads. Namely, for $\xi = 1/2$ the
Lorentz matrix (\ref{Lorentz}) transforms our tetrad to that of Pomeransky
and Khriplovich, and for $\xi = 1$ we obtain the tetrad of Landau and
Lifshitz.

Under the Lorentz transformation $e'{}^\alpha_i = \Lambda^\alpha{}_\beta
e^\beta_i$, the connection changes from $\Gamma_{i\alpha}{}^\beta$ to the
new one: $\Gamma{}'{}_{i\alpha}{}^\beta = \Lambda^\alpha{}_\gamma
\Gamma_{i\delta}{}^\gamma(\Lambda^{-1})^\delta{}_\beta + \Lambda^\alpha
{}_\gamma\partial_i(\Lambda^{-1})^\gamma{}_\beta$. Specifically, for the
weak gravitational field of a slowly rotating source (\ref{LT}) we find
\begin{eqnarray}
\Gamma{}'_{i\widehat{a}}{\,}^{\widehat{0}} &=& {\frac {GM\,x_a}{c^2r^3}}
\,e'_i{}^{\widehat{0}} - {\frac {3x_{(a}K_{b)}}{cr^2}}\,e'_i{}^{\widehat{b}}
+ \xi\,{\frac {\epsilon_{abc}\omega^b}{c^2}}\left( -\,\delta^c_d + {\frac
{3x^cx_d}{r^2}}\right)e'_i{}^{\widehat{d}},\label{conn1}\\
\Gamma{}'_{i\widehat{b}}{\,}^{\widehat{a}} &=& {\frac 12}\,\epsilon_{abc}
\omega^d\left( -\,\delta^c_d + {\frac {3x^cx_d}{r^2}}\right)e'_i{}^{\widehat{0}}
+ {\frac {GM}{c^2r^3}}\left(x^a e'_i{}^{\widehat{b}}
- x^b e'_i{}^{\widehat{a}}\right).\label{conn2}
\end{eqnarray}
Recall that $\Gamma{}'_{i\widehat{0}}{\,}^{\widehat{a}} = c^2\delta^{ab}
\Gamma{}'_{i\widehat{b}}{\,}^{\widehat{0}}$. We can drop the primes now,
since the value of the $\xi$ parameter identifies the reference frame
anyway.

Substituting (\ref{conn1}) and (\ref{conn2}) into (\ref{OmegaG}), we obtain
the precession of the physical spin in the gravitational field of rotating
object:
\begin{equation}\label{OmegaGK}
{\bm \Omega} = \gamma\left\{{\frac G{c^2r^3}}\left[{\frac {3{\bm r}({\bm r}
\cdot{\bm J})}{r^2}}-{\bm J}\right] + {\frac {{\bm \rho}\times {\bm v}}{c^2}}\right\},
\end{equation}
where we denote
\begin{equation}\label{rho}
{\bm \rho} = {\frac {2\gamma + 1}{\gamma + 1}}\,{\frac {GM}{r^3}}\,{\bm r}
+ {\frac {\gamma}{\gamma + 1}}\,{\frac {3G}{c^2r^3}}\left[{\frac {{\bm r}}{r^2}}
\,({\bm r}\cdot({\bm J}\times{\bm v})) - {\frac {2\xi}3}\,{\bm J}\times{\bm v}
+ (2\xi - 1){\frac {({\bm r}\cdot{\bm v})}{r^2}}\,{\bm J}\times{\bm r}\right].
\end{equation}
Putting $\xi = 0$, thus specifying to the {\it Schwinger} tetrad, we find
that the classical formula (\ref{OmegaGK}) perfectly reproduces the quantum
result (\ref{OmegaCl}). The extra Lorentz factor is due to the fact that the
classical evolution of spin was measured by using the proper time $\tau$,
whereas the quantum evolution was analyzed by using the coordinate time $t$.
If we choose another tetrad by putting $\xi = 1/2$
in (\ref{rho}), the equation (\ref{OmegaGK}) yields the result by
Pomeransky and Khriplovich \cite{PK} and Dvornikov
\cite{Dvornikov}:
\begin{equation} \begin{array}{c}
\bm\Omega^{(PK)}%_{LT}
= \frac{G}{c^2r^3} \left[\frac{3(\bm
r\cdot\bm J)\bm r}{r^2} - \bm J\right] -\frac{\gamma}{\gamma+1}
\frac{G}{c^2r^3}\left\{\frac{3[\bm r\times\bm v](\bm J
\cdot[\bm r\times\bm v])}{r^2} + \bm v\times(\bm v\times\bm J)\right\}.
\end{array} \label{OmePKtr}
\end{equation}
This result evidently differs from the Eq. (\ref{OmegaCl}).

It is worthwhile to mention that our result
(\ref{OmegaGK})-(\ref{rho}) (together with its quantum
(\ref{OmegaCl}) counterpart) presented in Ref. \cite{preprintLT}
was confirmed in the recent paper \cite{barausse} (specifically, cf.
the eq. (6.19) therein). This is very
satisfactory, since the authors of \cite{barausse} worked in a
different framework developing the Hamiltonian theory of a
spinning particle in a curved spacetime. In Ref. \cite{barausse},
the first relativistic corrections were calculated, while our Eqs.
(\ref{OmegaGK})-(\ref{rho}) and the corresponding quantum equations
are the exact formulas suitable also for the discussion of an
ultrarelativistic spin-$1/2$ particle. It is also stressed in
\cite{barausse} that the results obtained are consistent, in the
test-particle limit, with the earlier analysis \cite{jena} of the
dynamics of two gravitationally interacting rotating extended bodies.

%%%%%%%%%%%%%%%%%%%%%%%%%%%%%%%%%%%%%%%%%%%%%%%%%
\section{Particle with spin in a rotating frame}\label{rot}
%%%%%%%%%%%%%%%%%%%%%%%%%%%%%%%%%%%%%%%%%%%%%%%%%

One can straightforwardly show that Eq. (\ref{OmegaG}) yields the correct
angular velocity of spin precession in a rotating frame. The spacetime
is flat in this case with the line element \cite{HN} given by
\begin{equation}
ds^2 = \left[c^2 - (\bm{\omega}\times\bm r)^2\right]dt^2 - 2(\bm{\omega}
\times\bm r)_adx^adt - (dx^a)^2.\label{rotm}
\end{equation}
We choose the Schwinger gauge for the tetrad, which is then described by
(\ref{coframe}) with (\ref{VWni}), where we have to put ${\bm a} = 0$ for
the pure rotation. Other tetrads are easily obtained with the help of the
Lorentz transformation (\ref{Lorentz}) which is much simpler now because
$V = W = 1$. One can verify that the choosing $\xi = 1/2$ we indeed obtain
a symmetric tetrad (which corresponds to the gauge of Pomeransky and
Khriplovich), whereas $\xi = 1$ yields the Landau-Lifshitz tetrad.

The corresponding family of the (transformed) connection reads
\begin{equation}
\Gamma{}'_{i\widehat{a}}{\,}^{\widehat{0}} = -\,{\frac \xi{c^2}}
\,\epsilon_{abc}\omega^c\,e'_i{}^{\widehat{b}},\qquad
\Gamma{}'_{i\widehat{b}}{\,}^{\widehat{a}} = -\,\epsilon_{bac}
\omega^c\,e'_i{}^{\widehat{0}}.\label{conHN}
\end{equation}
Substituting this into the eq. (\ref{OmegaG}), we find the precession
angular velocity
\begin{equation}
{\bm\Omega} = \gamma\left(- {\bm\omega} + {\frac {\xi\,\gamma}{\gamma + 1}}
\,{\frac {\bm{v}\times(\bm{v}\times\bm\omega)}{c^2}}\right).\label{OmegaHN}
\end{equation}
The overall Lorentz factor is again due to the use of the proper time
in the evolution equations.

The correct result for the Schwinger gauge (that is recovered for $\xi = 0$)
was first obtained in \cite{Gor,Mashhoon}. A transparent and simple explanation
of the dependence of the angular velocity of the spin precession on the
gauge of a tetrad, and thus of the additional terms which are present in
(\ref{OmegaHN}) in the symmetric gauge (for $\xi = 1/2$) and in the
Landau-Lifshitz gauge (for $\xi = 1$), was presented recently \cite{Warszawa}
on the basis of the Thomas precession.

The study of a rotating frame helps to clarify the difference between the
tetrad gauges. The line element (\ref{rotm}) describes the flat spacetime. Indeed,
we can bring the metric to the explicitly flat form by a coordinate transformation
that replaces $(t, x^a)$ with the new coordinates $(T, X^a)$ using the formulas
\begin{equation}
t = T,\qquad x^a = L^a{}_b\,X^b.\label{xX}
\end{equation}
Here the $3\times 3$ matrix
\begin{equation}
L^a{}_b = n^a\,n_b + \left(\delta^a_b - n^a\,n_b\right)\cos\varphi
+ \epsilon^a{}_{cb}\,n^c\,\sin\varphi\label{rotL}
\end{equation}
defines a rotation around the unit vector $n^a$ on an angle $\varphi(T) = \omega\,T$
with the constant angular velocity $\dot{\varphi} = \omega$, where $\omega_a =
\omega\,n_a$. Differentiating (\ref{xX}), we easily verify
\begin{equation}
dx^a - K^acdt = L^a{}_b\,dX^a.\label{dxdX}
\end{equation}
Accordingly, the transformation (\ref{xX}) brings the line element (\ref{rotm})
to $ds^2 = c^2dT^2 - \delta_{ab}\,dX^adX^b$ which is the flat Minkowski world
described in the Cartesian coordinates $(T, X^a)$.

In order to compare different tetrad gauges, let us consider the continuous
family that arises from the Lorentz transformation (\ref{Lorentz}).
Explicitly, the tetrad components read
\begin{equation}
e^\alpha_i = \left(\begin{array}{c|c} e^{\widehat 0}_0 = \lambda(1 - \xi K^2) &
e^{\widehat 0}_b = \lambda\xi K_b/c\\ \hline e^{\widehat a}_0 =
\lambda (\xi - 1) c K^a & e^{\widehat a}_b= \delta^a_b + (\lambda - 1)K^aK_b/K^2
\end{array}\right).\label{hR}
\end{equation}
Here, as before, $\lambda = 1/\sqrt{1 - \xi^2K^2}$. As we can easily verify, this
family indeed contains all the three main options: (i) for $\xi = 0$, the Schwinger
gauge is obtained, (ii) the Landau-Lifshitz gauge arises for $\xi = 1$, (iii) the
symmetric gauge is recovered for $\xi = 1/2$.

Let us now analyse a particle that is {\it at rest} with respect to a reference system
described by the tetrad (\ref{hR}). The tetrad components of the 4-velocity of such
a particle read $U^\alpha = \delta^\alpha_{\hat 0} = (1, {\bf 0})$. Respectively,
the world components of particle's 4-velocity read $dx^i/d\tau = U^i = e^i_\alpha
U^\alpha = e^i_{\hat 0}$. Explicitly we then find from the inverse of (\ref{hR})
\begin{equation}\label{dxdt}
{\frac {dt}{d\tau}} = \lambda,\qquad {\frac {dx^a}{d\tau}} = \lambda c (1 - \xi)K^a
\quad \Longrightarrow \quad {\frac {dx^a}{dt}} = (1 - \xi)\epsilon^{abc}\omega_bx_c.
\end{equation}
This is how particle's dynamics is described in the coordinates $(t, x^a)$ of a
homogeneously rotating flat world. But how does this motion look in terms of the
genuinely inertial coordinates? Substituting (\ref{xX}) into (\ref{dxdt}), we obtain
\begin{equation}\label{dxdt1}
{\frac {dX^a}{dT}} = -\,\xi\epsilon^{abc}\omega_bX_c.
\end{equation}
As we see, for $\xi = 0$ the particle is at rest in the inertial coordinates,
$X^a =$const. That is, a {\it particle which is at rest in the Schwinger reference
frame indeed does not physically move} in the Cartesian coordinates $(T, X^a)$.
However, when a particle is at rest with respect to the Landau-Lifshitz $\xi = 1$
tetrad or with respect to the symmetric $\xi = 1/2$ tetrad, it actually turns out
to be moving (rotating) in the inertial coordinates!

In this sense, the Schwinger gauge is physically distinguished as it qualifies for
defining an {\it almost inertial} reference frame. In all the other tetrads with
$\xi \neq 0$, the description of physical effects is ``spoiled" by non-inertiality.
Such a spoiling effect is manifested by the additional (last) term in (\ref{OmegaHN}),
for example.

Certainly, one should be careful when generalizing the above observation
to the case of the nontrivial gravitational field. Note that we use the
expression ``almost inertial" to distinguish the anholonomic Schwinger
tetrad from a truly inertial tetrad which is holonomic. In a curved spacetime
one cannot, as a matter of principle, separate the inertial effects from the
gravitational ones. However, we recall again that Lense and Thirring have
demonstrated, in the weak-field approximation, the similarity of particle's
motion in a gravitational field of a rotating source to its dynamics in a
noninertial frame rotating with the angular velocity (\ref{equiv}). Our analysis
thus demonstrates that the Schwinger choice clearly appears to be preferable.

%%%%%%%%%%%%%%%%%%%%%%%%%%%%%%%%%%%%%%%%%%%%%%%%%
\section{Discussion and conclusion}\label{discussion}
%%%%%%%%%%%%%%%%%%%%%%%%%%%%%%%%%%%%%%%%%%%%%%%%%

In this paper, we consider the quantum and classical dynamics of a particle
with spin in the gravitational field of a rotating source. Being primarily
interested in the dynamics of spin, we derive the quantum-mechanical and
semiclassical equations of motion of the spin of a Dirac particle from the
Foldy-Wouthuysen approach. We demonstrate that the precession of the quantum
spin is in a perfect agreement with the motion of the classical spin derived
within a general scheme of Sec.~\ref{CS}. The results obtained are compared
to the earlier computations for different tetrad gauges, and we show that the
precession effect depends crucially on the choice of a tetrad. At the same
time, we find a perfect consistency with the classical Mathisson-Papapetrou
approach by explicitly calculating the quantum and semiclassical expressions
for the spin-dependent force on a Dirac particle.

The Lense-Thirring effect or frame dragging is one of the most impressive
predictions of the general relativity. This effect is currently analyzed
in the Gravity Probe B experiment \cite{GP-B,GP-B_problems}.
However, relativistic corrections to the LT effect are not observable in
this experiment as well as in other experiments inside the solar system
\cite{solar}.

Nevertheless, it is necessary to take the relativistic corrections to the
LT precession into account for the investigation of physical phenomena in
the binary stars such as pulsar systems. In this case, both components of a
system undergo a mutual Lense-Thirring precession about the total angular
momentum $\bm J$. Since the spin precession effects are well observable
\cite{radiopulsars,binaries,relpulsar}, the use of the results obtained
in the present work may be helpful for the high-precision calculations of
spin dynamics in the binaries.

%%%%%%%%%%%%%%%%%%%%%%%%%%%%%%%%%%%%%%%%%%%%%%%%%
\section*{Acknowledgments}
%%%%%%%%%%%%%%%%%%%%%%%%%%%%%%%%%%%%%%%%%%%%%%%%%

We are grateful to Lewis Ryder and Gerhard Sch\"afer for the
useful discussions and correspodence.
This work was supported in part by the Belarusian Republican
Foundation for Fundamental Research (grant No. $\Phi$08D-001),
the program of collaboration BLTP/Belarus, the Deutsche
Forschungsgemeinschaft (grants No. HE 528/21-1 and No. 436 RUS
113/881/0), the Russian Foundation for Basic Research (grants No.
09-02-01149 and No. 09-01-12179), and the Russian Federation
Ministry of Education and Science (grant No. MIREA 2.2.2.2.6546).
%%%%%%%%%%%%%%%%%%%%%%%%%%%%%%%%%%%%%%%%%%%%%%%%%%

%\newpage


\begin{thebibliography}{99}

\bibitem{LT}
H. Thirring,
{\it \"Uber die Wirkung rotierender ferner Massen in der
Einsteinschen Gravitationstheorie,} Phys. Z. %eitschrift
{\bf 19} (1918) 33-39 [English translation: {\it On the effect of 
rotating distant masses in Einstein's theory of gravitation]},
Gen. Rel. Grav. {\bf 16} (1984) 712-725]; H. Thirring, {\it Berichtigung 
zu meiner Arbeit: "\"Uber die Wirkung rotierender ferner Massen in der 
Einsteinschen Gravitationstheorie"}, Phys. Z. {\bf 22} (1921) 29-30 
[English translation: {\it Correction to my paper: ``On the effect of 
rotating distant masses in Einstein's theory of gravitation"}, 
Gen. Rel. Grav. {\bf 16} (1984) 725-727];
J. Lense and H. Thirring,
{\it \"Uber den Einflu{\ss} der Eigenrotation der
Zentralk\"orper auf die Bewegung der Planeten und Monde nach der
Einsteinschen Gravitationstheorie}, Phys. Z. {\bf 19} (1918) 156-163
[English translation: {\it On the influence of the proper rotation
of central bodies on the motions of planets and moons according to 
Einstein's theory of gravitation}, Gen. Rel. Grav. {\bf 16} (1984)
727-750].

\bibitem{Schiff}
L.I. Schiff,
{\it On experimental tests of the general theory of relativity},
{Am. J. Phys.} \textbf{28} (1960) 340-343; L.I. Schiff, 
{\it Motion of gyroscope according to Einstein's theory of gravitation},
{Proc. Nat. Acad. Sci.} \textbf{46} (1960) 871-882; L.I. Schiff,
{\it Possible new experimental test of general relativity theory}, 
{Phys. Rev. Lett.} \textbf{4} (1960) 215-217.

\bibitem{Shiff2}
I.Yu. Kobzarev and V.I. Zakharov,
{\it Spin precession in a gravitational field},
Ann. Phys. (USA) {\bf 37} (1966) 1-6;
R.F. O'Connell,
{\it Schiff's proposed gyroscope experiment as a test of the
scalar-tensor theory of general relativity},
Phys. Rev. Lett. {\bf 20} (1968) 69-71;
D.C. Wilkins,
{\it General equation for the precession of a gyroscope},
Ann. Phys. (USA) {\bf 61} (1970) 277-293;
B. Mashhoon,
{\it Partics with spin in a gravitational field},
J. Math. Phys. {\bf 12} (1971) 1075-1077.

\bibitem{Mathisson}
M. Mathisson, {\it Neue Mechanik materieller Systeme},
{Acta Phys. Polon.} {\bf 6} (1937) 163-200.

\bibitem{Papapetrou}
A. Papapetrou, 
{\it Spinning test particles in general relativity. I}, 
Proc. Roy. Soc. Lond. A {\bf 209} (1951) 248-258.

\bibitem{corben}
H.C. Corben, {\it Classical and quantum theories of spinning particles}
(Holden-Day, Inc: San Francisco, 1968) 279 pp.;
B.M. Barker and R.F. O'Connell,
{\it The gravitational interaction: Spin, rotation, and quantum effects -
A review},
Gen. Rel. Grav. {\bf 11} (1979) 149-175.

\bibitem{PK}
A.A. Pomeransky and I.B. Khriplovich,
{\it Equations of motion of spinning relativistic particle in external 
fields}, Zh. Eksp. Teor. Fiz. {\bf 113} (1998) 1537-1557 [J.\ Exp.\ Theor.\
Phys. {\bf 86} (1998) 839-849].

\bibitem{Dvornikov}
M. Dvornikov,
{\it Neutrino spin oscillations in gravitational field},
Int. J. Mod. Phys. D \textbf{15} (2006) 1017-1033; hep-ph/0601095.

\bibitem{HN}
F.W. Hehl and W.T. Ni,
{\it Inertial effects of a Dirac particle},
Phys. Rev. D {\bf 42} (1990) 2045-2048.

\bibitem{LL}
L.D. Landau and E.M. Lifshitz, The Classical Theory of Fields, 4th
revised English edition (Butterworth-Heinemann, Oxford, 1980)
Sec. 88, p. 272.
%Pergamon Press, Oxford (1975).

\bibitem{Ob1}
Yu.N. Obukhov, {\it Spin, gravity, and inertia}, Phys. Rev. Lett. 
{\bf 86} (2001) 192-195. 
%arXiv:gr-qc/0012102.

\bibitem{Ob2}
Yu.N. Obukhov, {\it On gravitational interactions of fermions}, 
Fortschr. Phys. {\bf 50} (2002) 711-716.
%, arXiv:gr-qc/0112080.

\bibitem{JMP}
A.J. Silenko,
{\it Foldy-Wouthuysen transformation forrelativistic particles 
in external fields}, J. Math. Phys. {\bf 44} (2003) 2952-2966.

\bibitem{PRA}
A.J. Silenko,
{\it Foldy-Wouthyusen transformation and semiclassical limit 
for relativistic particles in strong external fields},
Phys. Rev. A \textbf{77} (2008) 012116 [7 pages].

\bibitem{PRD}
A.J. Silenko and O.V. Teryaev,
{\it Semiclassical limit for Dirac particles interacting with 
a gravitational field},
Phys. Rev. D {\bf 71} (2005) 064016 [8 pages].

\bibitem{wald}
R. Wald,
{\it Gravitational spin interaction},
Phys. Rev. D {\bf 6} (1972) 406-413;
B.M. Barker and R.F. O'Connell,
{\it The gravitational interactions: Spin, rotation, and
quantum effects - A review},
Gen. Rel. Grav. {\bf 11} (1979) 149-175.

\bibitem{wong}
S.K. Wong,
{\it Heisenberg equations of motion for spin-1/2 wave equation
in general relativity},
Int. J. Theor. Phys. {\bf 5} (1972) 221-230;
L. Kannenberg,
{\it Mean motion of Dirac electrons in a gravitational field},
Ann. Phys (USA) {\bf 103} (1977) 64-73;
J. Audretsch and D.J. Diestler,
{\it Trajectories and spin motion of massive spin-1/2 particles
in gravitational fields},
J. Phys. A: Math. and Gen. {\bf A14} (1981) 411-422.

\bibitem{Plyatsko}
R. Plyatsko, {\it Gravitational ultrarelativistic spin-orbit 
interaction and the weak equivalence principle},
Phys. Rev. D \textbf{58} (1998) 084031 [5 pages];
R. Plyatsko, {\it Fermi-transported spinor and Dirac 
equation in general relativity}, arXiv: gr-qc/0601111 (2006).

\bibitem{singh}
B. Mashhoon and D. Singh,
{\it Dynamics of extended spinning masses in a gravitational field},
Phys. Rev. D {\bf 74} (2006) [12 pages];
D. Singh,
{\it Perturbation method for classical spinning particle motion.
I. Kerr space-time},
Phys. Rev. D {\bf 78} (2008) [21 pages];
D. Singh,
{\it An analytic perturbation approach for classical spinning
particle dynamics},
Gen. Rel. Grav. {\bf 40} (2008) 1179-1192.

\bibitem{KO}
I.Yu. Kobzarev and L.B. Okun,
{\it Gravitational interaction of fermions}, 
Zh. Eksp. Teor. Fiz. {\bf 43} (1962) 1904-1909 
[Sov. Phys. JETP {\bf 16} (1963) 1343-1346].

\bibitem{T1}
O.V. Teryaev,
{\it Spin structure of nucleon and equivalence principle},
arXiv: hep-ph/9904376 (1999); O.V. Teryaev, 
{\it Sources of time reversal odd spin asymmetries in QCD},
%Czech. J. Phys. {\bf 53}, 47 (2003)
arXiv: hep-ph/0306301 (2003).

\bibitem{PRD2}
A.J. Silenko and O.V. Teryaev, 
{\it Equivalence principle and experimental tests of gravitational
spin effects}, Phys. Rev. D {\bf 76} (2007) 061101(R) [5 pages].

\bibitem{Warszawa}
A.J. Silenko,
{\it Classical and quantum spins in curved spacetimes},
Acta Phys. Polon. B Proc. Suppl. {\bf 1} (2008) 87-107.

\bibitem{Mashhoon2}
B. Mashhoon,
{\it Quantum theory in accelerated frames of reference}, 
Lect. Notes Phys. {\bf 702} (2006) 112-132.

\bibitem{BMT}
V. Bargmann, L. Michel, and V.L. Telegdi,
{\it Precession of the polarization
of particles moving in a homogeneous electromagnetic field},
Phys. Rev. Lett. {\bf 2} (1959) 435-436.

\bibitem{Thoms}
L.H. Thomas, {\it The motion of the spinning electron},
Nature (London) {\bf 117} (1926) 514; L.H.~Thomas, 
{\it The kinematics of an electron with an axis},
Phil. Mag. (ser. 7) {\bf 3} (1927) 1-22.

\bibitem{Schwinger}
J. Schwinger,
{\it Energy and momentum density in field theory},
Phys. Rev. {\bf 130} (1963) 800-805;
J. Schwinger,
{\it Quantized gravitational field},
Phys. Rev. {\bf 130} (1963) 1253-1258.

\bibitem{dirac}
P.A.M. Dirac,
{\it Interacting gravitational and spinor fields},
in: {\sl ``Recent developments in general relativity"}
(Pergamon Press, Oxford and PWN, Warsaw, 1962) pp. 191-200.

\bibitem{LLp}
Ref. \cite{LL}, Sec. 98.

\bibitem{Gor}
A. Gorbatsevich, Exp. Tech. Phys. {\bf 27}, 529 (1979).

\bibitem{Mashhoon}
B. Mashhoon,
{\it Neutron interferometry in a rotating frame of reference},
Phys. Rev. Lett. {\bf 61} (1988) 2639-2642.

\bibitem{GP-B}
I. Ciufolini and E.C. Pavlis,
{\it A confirmation of the general
relativistic prediction of the Lense-Thirring effect},
Nature \textbf{431} (2004) 958-960.

\bibitem{GP-B_problems}
C.W.F. Everitt et al.,
{\it Gravity Probe B data analysis status and potential for 
improved accuracy of scientific results},
Class. Quantum Grav. \textbf{25} (2008) 114002 (11 pages).

\bibitem{solar}
L. Iorio, {\it Testing frame-dragging with the Mars Global 
Surveyor spacecraft in the gravitational field of Mars}, 
in \emph{The measurement of gravitomagnetism: A
challenging enterprise}, ed. by L. Iorio (Nova publishers,
Hauppauge (NY), 2007), pp. 177-187;
L. Iorio and V. Lainey,
{\it The Lense-Thirring effect in the Jovian system
of the Galilean satellites and its measurability},
Int. J. Mod. Phys. D \textbf{14} (2005) 2039-2050.

\bibitem{radiopulsars}
R.D. Blandford,
{\it Lense-Thirring precession of radio pulsars},
J. Astrophys. Astron., \textbf{16} (1995) 191-206.

\bibitem{binaries}
L. Stella and M. Vietri,
{\it Lense-Thirring precession and quasi-periodic 
oscillations inlow-mass X-ray binaries},
Astrophys. J. \textbf{492} (1998) L59-L62.

\bibitem{relpulsar}
A. W. Hotan, M. Bailes, and S. M. Ord,
{\it Geodetic precession in PSR J1141-6545},
Astrophys. J. \textbf{624} (2005) 906-913.

\bibitem{preprintLT}
Yu.N. Obukhov, A.J. Silenko, and O.V. Teryaev,
{\it Classical and quantum equations of motion of spin
for particles in nonstatic spacetimes},
in: {\sl Proc. of XIII Internat. Conf. ``Selected
Problems of Modern Physics", Dubna, 23-27 June 2008},
Eds. B.M. Barbashov and S.M. Eliseev (Joint Inst. Nucl.
Res., JINR, Dubna, 2009) pp. 168-170;
Yu.N. Obukhov, A.J. Silenko, and O.V. Teryaev,
{\it Spin dynamics in gravitational fields of rotating
bodies and the equivalence principle},
arXiv: 0907.4367 (gr-qc).

\bibitem{barausse}
E. Barausse, E. Racine, and A. Buonanno,
{\it Hamiltonian of a spinning test-particle in curved spacetime},
arXiv: 0907.4745 (gr-qc).

\bibitem{jena}
J. Steinhoff, S. Hergt, and G. Sch\"afer,
{\it ADM canonical formalism for gravitating spinning objects},
Phys. Rev. D {\bf 77} (2008) 104018 [16 pages];
J. Steinhoff, S. Hergt, and G. Sch\"afer,
{\it Spin-squared Hamiltonian of next-to-leading order
gravitational interaction},
Phys. Rev. D {\bf 78} (2008) 101503 [5 pages].

\end{thebibliography}
\end{document}